\documentclass[fleqn]{annalen}
\usepackage{graphics}
\usepackage{graphicx}
\usepackage{epsf}
\begin{document}
%%%%%%%%%%%%%%%%%%%%%%%%%%%%%%%%%%%%%%%%%%%%%%%%%%%%%%%%%%%%%%%%%%%%%%%%%%%%%%
%%%%%%%% the following newcommands will be completed by the publisher %%%%%%%%
%%%%%%%%%%%%%%%%%%%%%%%%%%%%%%%%%%%%%%%%%%%%%%%%%%%%%%%%%%%%%%%%%%%%%%%%%%%%%%
\newcommand{\volume}{8}              %sets current volume,
\newcommand{\xyear}{1999}            %sets year in header
\newcommand{\issue}{5}               %sets current issue,
\newcommand{\recdate}{29 July 1999}  %sets received date,
\newcommand{\revdate}{dd.mm.yyyy}    %sets revised date,
\newcommand{\revnum}{0}              %number of revisions,
\newcommand{\accdate}{dd.mm.yyyy}    %sets accepted date,
\newcommand{\coeditor}{ue}           %sets (co)editor,
\newcommand{\firstpage}{507}         %first page number,
\newcommand{\lastpage}{510}          %last page number,
\setcounter{page}{\firstpage}        %sets page counter to first page number
%%%%%%%%%%%%%%%%%%%%%%%%%%%%%%%%%%%%%%%%%%%%%%%%%%%%%%%%%%%%%%%%%%%%%%%%%%%%%%
%%%%%%%%%%%%%%%%%%%%%%%%%%%%%%%%%%%%%%%%%%%%%%%%%%%%%%%%%%%%%%%%%%%%%%%%%%%%%%
%%%%%%%%%%%%%%%%%% please give up to three keywords here %%%%%%%%%%%%%%%%%%%%%
%%%%%%%%%%%%%%%%%%%%%%%%%%%%%%%%%%%%%%%%%%%%%%%%%%%%%%%%%%%%%%%%%%%%%%%%%%%%%%
       \newcommand{\keywords}{localization, scaling, multifractality}
%%%%%%%%%%%%%%%%%%%%%%%%%%%%%%%%%%%%%%%%%%%%%%%%%%%%%%%%%%%%%%%%%%%%%%%%%%%%%%
%%%%%%%%%%%%%%%% please give up to three PACS numbers here %%%%%%%%%%%%%%%%%%%
%%%%%%%%%%%%%%%%%%%%%%%%%%%%%%%%%%%%%%%%%%%%%%%%%%%%%%%%%%%%%%%%%%%%%%%%%%%%%%
\newcommand{\PACS}{72.15.Rn, 72.20.Ee, 73.20.Fz}
\newcommand{\shorttitle}{
L. S. Levitov, Critical Hamiltonians with long range hopping}
%% sets the header on oddpage
\title{Critical Hamiltonians with long range hopping}
\author{L. S. Levitov}
\newcommand{\address}
  {Physics Department, Center for Materials Sciences \& Engineering,
  Massachusetts Institute of Technology, 77 Massachusetts Ave.,
  Cambridge, MA 02139, USA}
\newcommand{\email}{\tt levitov@mit.edu}
\maketitle
%%%%%%%%%%%%%%%%%%%%%%%%%%%%%%%%%%%%%%%%%%%%%%%%%%%%%%%%%%%%%%%%%%%%%%%%%%%%%
\begin{abstract}
  Critical states are studied by a real space RG in the problem
with strong diagonal disorder and long range power law hopping.
The RG flow of the distribution of coupling parameters is
characterized by a family of non-trivial fix points. We consider
the RG flow of the distribution of participation ratios of 
eigenstates. Scaling of participation ratios is sensitive to
the nature of the RG fix point. 
For some fix points, scaling of participation ratios is 
characterized by a {\it distribution} of exponents, rather than
by a single exponent.

The RG method can be generalized to treat certain fermionic
Hamiltonians with disorder and long range hopping. We derive the
RG for a model of interacting two-level systems. Besides
couplings, in this problem the RG includes the density of
states. The density of states is renormalized so that it
develops a singularity near zero energy.

\end{abstract}
%%%%%%%%%%%%%%%%%%%%%%%%%%%%%%%%%%%%%%%%%%%%%%%%%%%%%%%%%%%%%%%%%%%%%%%%%%%%%
\vskip5mm
\noindent
Conventionally, localization is studied for Hamiltonians with
matrix elements rapidly decreasing as a function of the distance.
For instance, the canonical Anderson model is defined on a
lattice so that disorder is diagonal and hopping occurs between
neighboring sites. In this model the localization transition is
reached by reducing the amplitude of hopping $V$ below the
threshold $V_c$ approximately given by the disorder bandwidth
$W$ divided by the number of nearest neighbors.

The situation is completely different for the problem with long
range hopping\cite{Anderson}. In a $d-$dimensional space, for
hopping amplitude falling as some power of the distance,
$V(r-r')\sim |r-r'|^{-\alpha}$, all states are extended for
$\alpha\le d$ for both strong and weak disorder. At the same
time, for $\alpha>d$ and strong disorder the states are
localized.

Thus, one can study localization transition by
varying the exponent $\alpha$ at fixed disorder
strength \cite{LL,LL1,MirlinFyodorov,Kravtsov}.
This transition is more tractable than the
conventional one because of the availability of a small
parameter given by the ratio of hopping to disorder
strength \cite{LL,LL1}. At the transition line $\alpha=d$, a real space
renormalization group (RG) can be constructed for the
distribution of couplings. The RG flow obeys certain
conservation laws, and, depending on the details of the
Hamiltonian, the flow may or may not have a nontrivial fix
point.

Recently, this problem reappeared in the context
of quantum chaos in Kepler billiards\cite{LL1}. 
The RG approach was used in \cite{LL1} to treat scaling of wave
functions at the transition. In Sec.~\ref{Scaling} below the
RG scheme is applied to study the distribution of participation
ratios. We find that, depending on the type of the fix point,
participation ratios are either characterized by a single
scaling exponent, or by a {\it distribution} of scaling exponents.

One can extend the RG method \cite{LL,LL1} to certain many-body
systems with disorder and long range interactions. One
interesting example is the quantum problem of two
level systems with long range coupling \cite{Yu,EfShk,BhattLee}.
In Sec.~\ref{fermions} we
develop an RG for a $d=3$ problem with couplings falling
as a cube of the distance. From the RG flow derived for the
distribution of the wavefunctions and for the
density of states we find that:\\
 (i) Wavefunctions of all excitations are delocalized;\\
 (ii) Under RG of the density of states a
singularity develops near zero energy due to states with
small energy condensing into a peak. We discuss the
relation of this new state of interacting two level systems to
the Efros-Shklovskii Coulomb gap state \cite{EfShk}.

\section{Renormalization group}
   \label{RG}
We consider an ensemble of Hamiltonians on
a $d-$dimensional lattice:
  \begin{equation}
\label{Hamiltonian}
E \psi_{\bf r}= E_{\bf r} \psi_{\bf r}+\sum\limits_{\bf r'\ne r}V_{\bf r, r'} \psi_{\bf r'}\ ,
\qquad {\rm where}\qquad
V_{\bf r, r'}={\vec a_{\bf r}\cdot\vec a_{\bf r'}\over |{\bf r}-{\bf r'}|^d}\ .
  \end{equation}
Here $E_{\bf r}$ are random numbers, and $\vec a_{\bf r}$ are
random $n-$component vectors. The distribution of $E_{\bf r}$ is
uniform with density $\nu$: $dN/dE=\nu$ for $|E|<1/2\nu$; $0$
for $|E|>1/2\nu$. We assume that the distribution of $\vec
a_{\bf r}$, $dP=f(\vec a)d^na$, is such that all moments of
$f(\vec a)$ exist. Otherwise, the exact form of $f(\vec a)$ is
unimportant.

The coupling strength is characterized by dimensionless
parameter $\lambda=\langle a^2_{\bf r}\rangle n \nu$,
where $n$ is the density of lattice sites. In the weak coupling
regime, $\lambda\ll 1$, a real space RG
flow of the Hamiltonians of the form (\ref{Hamiltonian}) can be derived.
This is done by considering resonance pairs and
arguing that they correctly account for the flow of the coupling
distribution to the leading logarithmic order (see  \cite{LL,LL1}).

The RG flow preserves the form of the hopping amplitudes
$V_{\bf r, r'}$ in (\ref{Hamiltonian}), and modifies the distribution
$f(\vec a)$. The distribution of energies $E_{\bf r}$ is
renormalized near the tails \cite{Burin}, but remains constant in
the middle part. In the middle of the spectrum,
 $|E_{\bf r}|\ll\nu^{-1}$, one can treat $E_{\bf r}$ as
quasirandom numbers with fixed uniform distribution.

To introduce some objects appearing in the RG equation, let us
explicitly diagonalize the problem of a resonance between two
sites ${\bf r_1}$ and ${\bf r_2}$:
  \begin{equation} \label{(3.4)}
E \psi_{1}=E_{1}\psi_{1}+ V_{\bf r_1, r_2} \psi_{2}\ , \qquad
E \psi_{2}=E_{2}\psi_{2}+ V_{\bf r_2,r_1} \psi_{1}\ .
  \end{equation}
Two eigenstates $\psi^{+}$ and $\psi^{-}$ are defined by
  \begin{equation}
\psi^{+} = \cos \theta\ \psi_{1} + \sin \theta\ \psi_{2}, \qquad
\psi^{-} = -\sin \theta\ \psi_{1} + \cos \theta\ \psi_{2}\ ,
  \end{equation}
with
  \begin{equation} \label{(3.5)}
\tan\theta = \sqrt{\tau^2+1} - \tau\ , \qquad
\tau = \frac{E_{1}-E_{2}}{2V_{\bf r_1, r_2}}\ .
  \end{equation}
The energies of the states (\ref{(3.5)}) are
  \begin{equation} \label{(3.6)}
E_{\pm} = \frac{1}{2}\left( E_{1}+E_{2}
\pm \sqrt{(E_{1}-E_{2})^2 +4V^2_{\bf r_1,\bf r_2} }\right)
  \end{equation}
The transformation rule for the vectors $\vec a_{1,2}$ follows from the
relation
  \begin{equation} \label{(3.7)}
\vec a_{1} \psi_{1}+\vec a_{2} \psi_{2} =
\vec a_{+}\psi^{+} + \vec a_{-}\psi^{-}\ .
  \end{equation}
One obtains
  \begin{equation}
\vec a_{+}= \cos \theta\ \vec a_{1} + \sin \theta\ \vec a_{2}\ ,\qquad
\vec a_{-}= -\sin \theta\ \vec a_{1} + \cos \theta\ \vec a_{2}
  \label{(3.8)}
  \end{equation}
The role of transformed $\vec a_{\pm}$ is that they determine
hopping matrix elements
  $V_{\bf r, r_\pm}= \vec a_{\bf r}\cdot\vec a_{\pm}/|{\bf r}-{\bf r_\pm} |^d$
for all remote sites
  $|{\bf r}-{\bf r_\pm} |\gg|{\bf r_1}-{\bf r_2}|$,
where ${\bf r_\pm}=({\bf r_1}+{\bf r_2})/2$.

Subsequently diagonalizing pair
resonances at all distances, from small to large, one
derives \cite{LL} an RG equation for the distribution of $\vec
a_{\bf r}$'s. It has a form of a first order differential
equation with respect to the RG time $t=\ln \left(L\right)$
with an integral operator on the right hand side:
  \begin{eqnarray}
\frac{\partial}{\partial t}f\left({a}\right) &=& 2A_{\rm d}\nu n \int d\tau
d^n{a}_{1}
d^n{a}_{2}
\ |\vec a_1\cdot\vec a_2|\
f\left({a}_{1}\right) f\left({a}_{2}\right)
  \cr
&\times &
\Bigl[\delta\left({a}-
{a}_{+}\right)+\delta\left({a}-{a}_{-}\right)-
\delta\left({a}-{a}_{1}\right)-\delta\left({a}-
{a}_{2}\right)\Bigr]\ .
\label{(3.13)}
  \end{eqnarray}
Here $A_{\rm d}$, the area of a $d-$dimensional sphere, arises from 
$dV=A_{\rm d}r^{d-1}dr$.

The flow of the distribution $f\left({a}, t\right)$ obtained by
solving (\ref{(3.13)}) has a number of properties similar to
that of Boltzmann kinetic equation. The reason is that in
considering only pair resonances and ignoring correlations
between resonances at different scales one makes an approximation of
the same nature as the pair collision approximation in Boltzmann
theory of weakly nonideal gases.

Two properties of the flow (\ref{(3.13)}) are of interest. First, all second moments
  \begin{equation}
G_{\alpha\beta}=\
\langle a_\alpha a_\beta\rangle\ =\ \int f(a) a_\alpha a_\beta d^na
  \end{equation}
are conserved. Second, one can define entropy
  \begin{equation}\label{entropy}
H[f]\ =\ -\int \ln\left( |a|f(a)\right) f(a) d^na
  \end{equation}
and, for the flow (\ref{(3.13)}) modified by
$|\vec a_1\cdot\vec a_2|\to |\vec a_1||\vec a_2|$,
prove an H-theorem, $\partial H/\partial t\ge 0$.

These results can be used to fully characterize the fix points of the
flow (\ref{(3.13)}). By maximizing the
entropy (\ref{entropy}) for given moments $G_{\alpha\beta}$ one
obtains the family
  \begin{equation}\label{fix-point}
f_G(a) =\ \left(A/|a|\right)\exp\left(-a_\alpha G_{\alpha\beta}a_\beta\right)\ .
  \end{equation}
The normalization constant $A$ depends on the matrix $G$. 

%(By a simulation we checked that all conclusions about modified flow
%are qualitatively and even quantitatively applicable for the flow (\ref{(3.13)})).

It is interesting that the distributions (\ref{fix-point}) are
normalizable only when the number of components $n$ is greater
than one (and matrix $G_{\alpha\beta}$ is nondegenerate). Because of that,
the flow (\ref{(3.13)}) has very different properties for $n=1$ and $n>1$.

In the $n>1$ case, for any initial conditions $f(a)_{t=0}$, the
distribution $f(a)$ evolves to one of stationary distributions
(\ref{fix-point}). On the other hand, for $n=1$ possible fix point
distributions allowed by H-theorem are non-normalizable.
Therefore, since $\int f(a)_{t=0}da =1$, the flow (\ref{(3.13)})
does not have a fix point.

%%%%%%%%%%%%%%%%%%%%%%%%%%%%%%%%%%%%%%%%%%%%%%%%%%%%%%%%%%%%%%%%%%%%

\begin{figure}
\centerline{\resizebox{11cm}{7.5cm}{\includegraphics{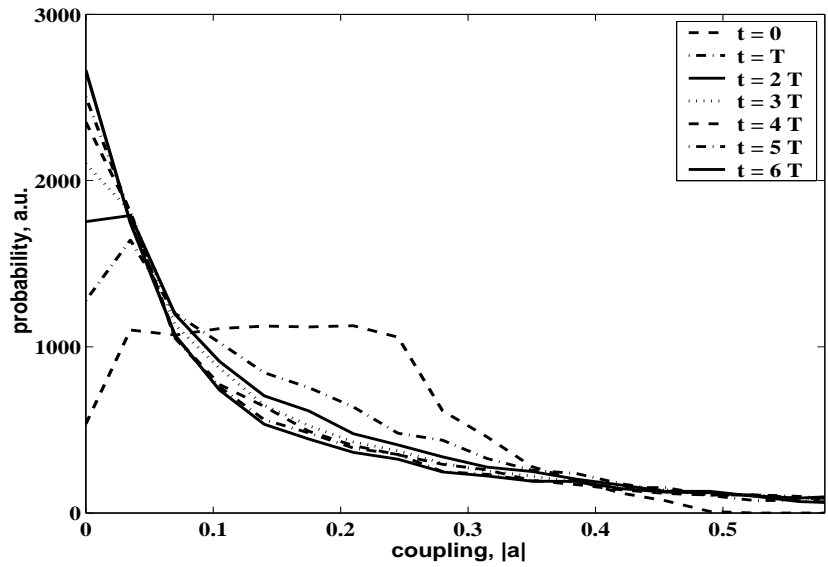}}}
%\centerline{\includegraphics[4.25cm][4.25cm]{fig1.eps}}
 \caption[]{The flow (\ref{(3.13)}) for $n=2$. Initial distribution is uniform
    in the disk $a^2_1+a^2_2<r^2_0$, and hence $f(a)$ is circular symmetric.
    At large times the distribution approaches a fix point of the form (\ref{fix-point})
    Parameters used in this simulation are: $\nu=1$, $r_0=0.6$, $T=4$.
    Dimensionless coupling $\lambda=0.2$ for these parameters.
    }
 \label{fig1}
\end{figure}
%%%%%%%%%%%%%%%%%%%%%%%%%%%%%%%%%%%%%%%%%%%%%%%%%%%%%%%%%%%%%%%%%%%%

\begin{figure}
%\vspace{-0.2cm}
\centerline{\resizebox{11cm}{7.5cm}{\includegraphics{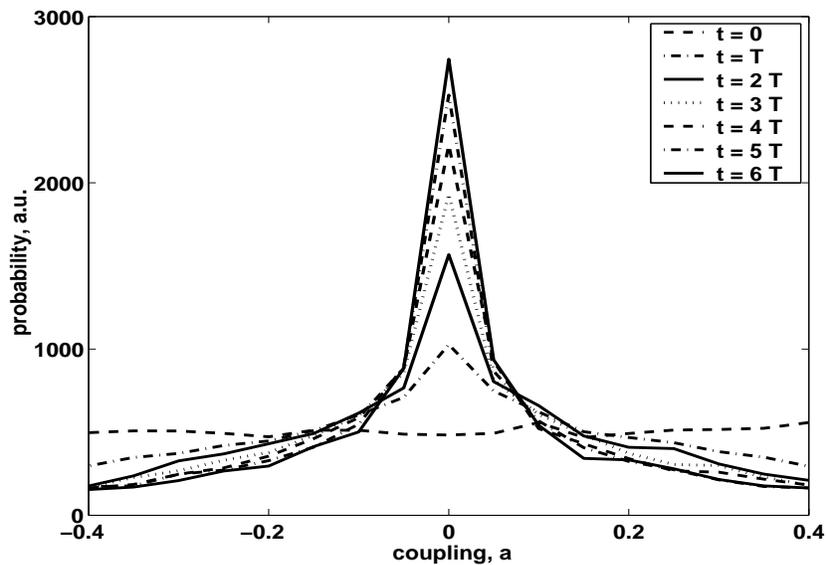}}}
 \caption[]{The flow (\ref{(3.13)}) for $n=1$. Initial distribution is uniform
    in the interval $|a|<r_0$.
    At large times the distribution looks more and more like $(A/|a|) e^{-ga^2}$
    with time-dependent $A$ and $g$. Note the divergence developing near $a=0$ and
    in the tails.
    Parameters used in this simulation are: $\nu=1$, $r_0=0.5$, $T=4$.
    Dimensionless coupling $\lambda=0.08$.
    }
 \label{fig2}
\end{figure}

These two qualitatively different types of flow are displayed in
Figures~\ref{fig1},\ref{fig2}. The simulation was performed by
using $N=10^4$ states characterized by random energies and
coupling vectors. The energy distribution was uniform with
$\nu=1$ and did not evolve. The distribution of $\vec a$ evolved
to a fix point of the form (\ref{fix-point}) for $n=2$ and diverged for $n=1$.

%%%%%%%%%%%%%%%%%%%%%%%%%%%%%%%%%%%%%%%%%%%%%%%%%%%%%%%%%%%%%%%%%%%%%5

%%%%%%%%%%%%%%%%%%%%%%%%%%%%%%%%%%%%%%%%%%%%%%%%%%%%%%%%%%%%%%%%%%

\section{Scaling of wavefunctions}
  \label{Scaling}
The RG hierarchy of states constructed as
resonances on different scales implies fractal structure the
wavefunctions. To understand scaling of
eigenstates, one can extend RG by including in it the distribution of
wavefunction amplitudes.

Below we consider the distribution of participation ratios of
the states. For a normalized state $\psi^{(i)}_{\bf r}$ its
participation ratio $p_4$ is defined as
$p_4=\sum\limits_{\bf r} |\psi^{(i)}_{\bf r}|^{4}$
In the localization problem the participation ratios are used as
a measure of the degree of localization. The scaling exponent
$\mu$ of participation ratios defined as $p\sim L^{-\mu}$, where
$L$ is the system size, characterizes possible universality
classes. In the localized regime $\mu=0$, in the delocalized
regime $\mu=d$, and at the critical point $0<\mu<d$.

It is straightforward to put the distribution of participation ratios
in the RG scheme. For that one has to consider the
flow of the distribution $f(a,p)$. Following the discussion of Sec.~\ref{RG},
one finds the participation ratios
$p_\pm$ of the eigenstates
of the resonance
pair (\ref{(3.4)}) in terms of
$p_{1}$ and $p_{2}$. Since the states $1$ and
$2$ do not overlap, one simply obtains:
  \begin{equation}
p_{+} =\ \cos^{4}\theta\  p_{1} + \sin^{4}\theta\  p_{2},\qquad
\qquad p_{-} =\ \sin^{4}\theta\  p_{1} + \cos^{4}\theta\  p_{2}.
    \label{(7.3)}
   \end{equation}
This change of the participation ratios must be considered
together with the transformation of the parameters $a_1$, $a_2$
given by Eq.~(\ref{(3.8)}). The resulting RG equation reads:
  \begin{eqnarray}
\frac{\partial}{\partial t}f\left({a},p\right)
&=&
2A_{\rm d}\nu n \int d\tau d^na_1 d^na_2
dp_{1}dp_{2}\ |\vec a_1\cdot\vec a_2|\
f\left({a}_{1},p_{1}\right)f\left({a}_{2},p_{2}\right) \cr
&\times& \big[\delta\left({a}-{a}_{+}\right)\delta\left(p-p_{+}\right)
+\delta\left({a}-{a}_{-}\right)\delta\left(p-p_{-}\right)  \cr
& &
- \delta\left({a}-{a}_{1}\right)\delta\left(p-p_{1}\right)
-\delta\left({a}-{a}_{2}\right)\delta\left(p-p_{2}\right)\big]\  .\qquad
    \label{(7.4)}
   \end{eqnarray}
Generally, one cannot factor the distribution
$f(a,p)$, because
the flow (\ref{(7.4)}) generates nontrivial correlations between
$\vec a$ and $p$.

%%%%%%%%%%%%%%%%%%%%%%%%%%%%%%%%%%%%%%%%%%%%%%%%%%%%%%%%%%%%%%%%%%%%
\begin{figure}
\centerline{\resizebox{10cm}{7cm}{\includegraphics{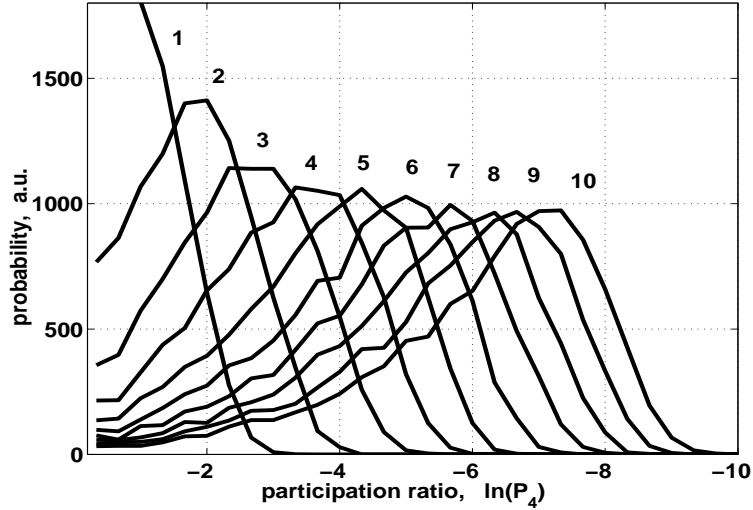}}}
 \caption[]{The flow of participation ratios for $n=2$. 
    All parameters are the same as in \protect{Figure~\ref{fig1}}. 
    Distributions are numbered according to the RG times $t_m=2mT$, $1\le m\le10$. 
    }
 \label{fig3}
\end{figure}
%%%%%%%%%%%%%%%%%%%%%%%%%%%%%%%%%%%%%%%%%%%%%%%%%%%%%%%%%%%%%%%%%%

It is of interest to consider the flow (\ref{(7.4)}) in the scaling
limit, when the distribution of $\vec a$ given by $\int f(a,p)
dp$ evolves to the fix points discussed in Sec.~\ref{RG}.
We studied numerically evolution of the participation ratios
distribution $\int f(a,p)d^na$. 

\begin{figure}
\centerline{\resizebox{10cm}{7cm}{\includegraphics{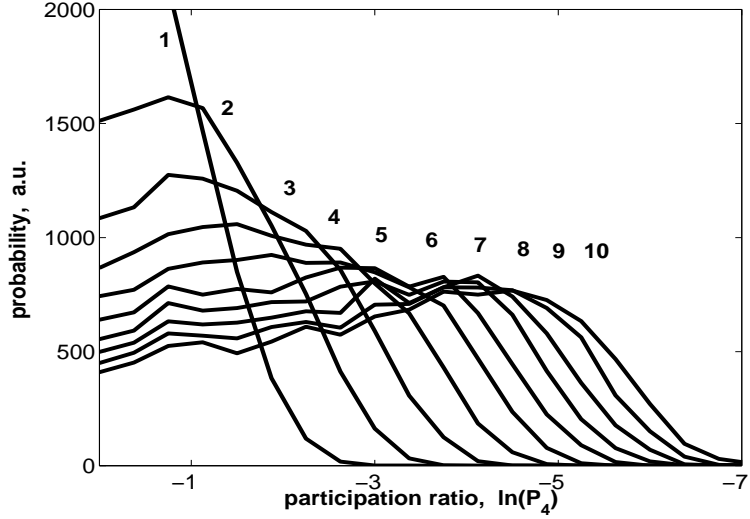}}}
 \caption[]{The flow of participation ratios for $n=1$ corresponding
    to the parameters used in \protect{Figure~\ref{fig2}}. 
    The RG times $t_m=2mT$, $1\le m\le10$. 
    }
 \label{fig4}
\end{figure}

Because of the expected power law scaling $p\sim
L^{-\mu}=\exp(-\mu t)$ it is natural to consider the
distribution of $\ln p$. The results of the simulation for $n=2$
and $n=1$ are shown in Figures~\ref{fig3},\ref{fig4}. Parameters
used are the same as in Figures~\ref{fig1},\ref{fig2}.

The results for $n=2$
and $n=1$ are quite different. For $n=2$, the distribution of $\ln p$ forms a peak
of slowly varying width. The center of the distribution moves linearly with the RG time $t$,
which corresponds to a single scaling exponent $\mu$ given by the peak velocity. 
In contrast, for $n=1$, the left edge of the distribution is
fixed at $\ln p=0$. After rescaling, the
distributions at different $t$ collapse on one another. This
means that participation ratios are characterized by a
{\it continuous spectrum} of scaling exponents $\mu$, rather than by a
single exponent.

The values of $\mu$ are not universal because of
a many-parameter family of fix points parameterized by
$G_{\alpha\beta}$ as discussed in Sec.~\ref{RG}. Even for
spherically symmetric distributions with
$G_{\alpha\beta}=G\delta_{\alpha\beta}$, the distribution
depends on the dimensionless parameter $\lambda=Gn\nu$. The
dependence on $\lambda $ however is quite simple. Numerically,
we find that doubling $\lambda$ doubles  velocity
of the center of the distribution of $\ln p$, without changing of
its overall form. This implies that scaling exponents $\mu$ can
be written as $\gamma\lambda$, where $\gamma\simeq1$ and has a universal distribution.

This estimate is in agreement with the result 
for the mean participation ratio \cite{LL1}, where it was
found that $\langle p\rangle$ scales as $L^{-\gamma\lambda}$
with $\gamma\simeq1$.

%%%%%%%%%%%%%%%%%%%%%%%%%%%%%%%%%%%%%%%%%%%%%%%%%%%%%%%%%%%%%%%%%%

%%%%%%%%%%%%%%%%%%%%%%%%%%%%%%%%%%%%%%%%%%%%%%%%%%%%%%%%%%%%%%%%%%

\section{RG for fermionic Hamiltonians}
  \label{fermions}
In some cases, the RG approach outlined above can be extended to
many-body problems \cite{BhattLee}. Here we consider a
Hamiltonian for interacting two-level systems in $d=3$:
   \begin{equation}
\label{TLS}
{\cal H}=\sum\limits_i
{\epsilon}_i{\sigma}^z_i+{1\over2}
\sum\limits_{i\ne j}
r_{ij}^{-3}V^{ab}_{ij}
{\sigma}^a_i{\sigma}^b_j
   \end{equation}
Here ${\sigma}^a_i$ are pseudospin Pauli matrices ($a,b=x,y,z$), and
$r_{ij}=|r_i-r_j|$ are distances between two-level systems
randomly distributed in space with concentration $n$. Both
${\epsilon}_i$ and $V^{ab}_{ij}$ are random
uncorrelated numbers.

The problem (\ref{TLS}) describes several physical situations, such as
the dynamics of excitations in two-level systems in glasses
interacting via elastic strain \cite{Yu}, or dipole excitations in the
Efros-Shklovskii model of a disordered Mott insulator \cite{EfShk}.
In the latter problem, the conventional approach is to consider
only $\sigma^z_i\sigma^z_j$ couplings in (\ref{TLS}). In this
case excitations are strictly localized, and the only effect of
interaction is the appearance of a logarithmic gap in the
density of states at low energies.

Here we focus on the opposite limit, and consider the effect of
$x-x$, $y-y$, and $x-y$ couplings, ignoring other couplings.
We assume that the interaction is weak,
$\lambda={\langle}|V^{ab}_{ij}|{\rangle}n/{\langle}|\epsilon_i|{\rangle}\ll 1$,
and develop an RG scheme using this weak coupling.
The $XY$ approximation is frequently used for studying spin
waves in spin systems. Although it does not seem to be
controlled by any small parameter, one should mention that the
terms removed have zero matrix elements between different states
with close energy. Therefore, these terms cannot produce
resonances (which are responsible for delocalization), so they
are unlikely to be important as long as delocalization of
states is concerned.

Fermionic model is obtained by replacing Pauli
matrices in (\ref{TLS}) by Fermi operators: ${\sigma}^x
\rightarrow a+a^+,{\ } {\sigma}^y \rightarrow i(a-a^+), {\ }
{\sigma}^z \rightarrow a^+a-aa^+$. Our main reason to employ
this spin--pseudofermion transformation is that it gives an
exact result for a pair of interacting spins, whereas only
interacting pairs contribute to the leading logarithmic order in
the RG. Thus we obtain
  \begin{equation}
\label{H-fermi}
{\cal H}=\sum\limits_i
{\epsilon}_i(a^+_ia_i-a_ia^+_i)+{1\over2}\sum\limits_{i\ne j}
r_{ij}^{-3}( u_{ij}
a^+_ia_j+{ v}_{ij}a_ia_j+H.c.)\ ,
  \end{equation}
where $ u_{ij}=V^{xx}_{ij}+V^{yy}_{ij}+ i(V^{xy}_{ij}-V^{yx}_{ij})$,
$ v_{ij}=V^{xx}_{ij}-V^{yy}_{ij}-i(V^{xy}_{ij}+V^{yx}_{ij})$.
Distributions of real $\epsilon_i$ and complex $ u_{ij}$,
$ v_{ij}$ are assumed to be uncorrelated.
We denote them by $\nu(\epsilon)$ and
$f( u, v)$, respectively: $dP=\nu(\epsilon)d\epsilon$
and $dP=f({ u},{ v})d^2{ u}d^2{ v}$.
The Hamiltonian (\ref{H-fermi}) is bilinear in $a$, $a^+$ and
can be treated perturbatively by an RG in the weak coupling
regime ${\langle}|u|{\rangle}{\nu}n{\ll}1$,
${\langle}|v|{\rangle}{\nu}n{\ll}1$.

The RG scheme is based on resonance pairs of pseudofermions:
  \begin{equation}
  \label{f-pair}
{\cal H}={\epsilon}_1(a^+_1a_1-a_1a^+_1)+{\epsilon}_2(a^+_2a_2-a_2a^+_2)+
r_{12}^{-3}({ u} a^+_1a_2+{ v}a_1a_2+H.c.)
  \end{equation}
The Hamiltonian (\ref{f-pair}) can be diagonalized by a two step Bogoliubov transformation:
(1) $(\tilde a_1,\tilde a_2)
=R_{\alpha}  (a_1,a_2)$; (2) $({a'}^+_1,a'_2)=R_{\beta} (\tilde a_1^+,\tilde a_2)$,
where
  \begin{equation}
R_{\alpha}=\left(\matrix{\cos\alpha &z_1\sin\alpha \cr
                          -\sin\alpha &z_1\cos\alpha \cr}\right),{\ \ }
R_{\beta}=\left(\matrix{\cos\beta &z_2\sin\beta \cr
                          -\sin\beta &z_2\cos\beta \cr}\right)
\ ,
  \end{equation}
  \begin{equation}
\tan 2\alpha ={{r_{12}^{-3}{ u}} \over \epsilon_1-\epsilon_2},
{\ \ }z_1={{ u} \over {|{ u}|}},{\ \ }
\tan 2\beta =-{{r_{12}^{-3}{ v}} \over \epsilon_1+\epsilon_2},{\ \ }
z_2={{ v} \over {|{ v}|}}\ .
  \end{equation}
The resulting diagonal Hamiltonian is
${\cal H}={\epsilon}'_1(a'^+_1a'_1-a'_1a'^+_1)+
{\epsilon}'_2(a'^+_2a'_2-a'_2a'^+_2)$, where
  \begin{equation}
({\epsilon}'_1-{\epsilon}'_2)^2=({\epsilon}_1-{\epsilon}_2)^2+
|{ u}r_{12}^{-3}|^2,
\hskip 5mm
({\epsilon}'_1+{\epsilon}'_2)^2=({\epsilon}_1+{\epsilon}_2)^2+
|{ v}r_{12}^{-3}|^2.
  \end{equation}
For another two-level system (say, described by $b$ and $b^+$)
the interaction with $a_1$ and $a_2$ is given by
  \begin{equation}
{\cal H}_{\rm int}=r_{1b}^{-3}({ u}_1a_1^+b+{ v}_1a_1b+H.c.)+
r_{2b}^{-3}({ u}_2a_2^+b+{ v}_2a_2b+H.c.)
  \end{equation}
In accordance with the picture of resonance pairs we can
consider only the case
$r_{1b}{\gg}r_{12}$,{\ } $r_{2b}{\gg}r_{12}$.
Therefore, $r_{1b}$ is almost equal
to $r_{2b}$ and hence the
two Bogoliubov transformations act on ${ u}_{1,2}$ and ${ v}_{1,2}$
as
  \begin{eqnarray}
(\tilde { u}_1,{\tilde  u}_2)=R_{\alpha}({ u}_1,
{ u}_2),\qquad
(\tilde { v}_1,\tilde { v}_2)=R_{\alpha}^*({ v}_1,{ v}_2);
\nonumber \\
  ({ u}'_1,{ v}'_2)=R_{\beta}^*(\tilde { u}_1,
\tilde { v}_2),\qquad
({ v}'_1,{ u}'_2)=R_{\beta}(\tilde { v}_1,\tilde { u}_2)\ ,
  \end{eqnarray}
where $*$ stands for complex conjugate. Similarly to how it was done
in Sec.s~\ref{RG},\ref{Scaling}, these expressions can be
used to derive an RG flow for distributions:
  \begin{equation}
  \label{f-rg}
{{\partial}\over{{\partial}{ t}}}g(\xi)=
2{\pi}ne^{3{ t}}{\langle}{\delta}(\xi-\xi'_1)+
{\delta}(\xi-\xi'_2)
-{\delta}(\xi-\xi_1)-{\delta}(\xi-\xi_2){\rangle}_{{ u},{ v},\xi_1,\xi_2}\ ,
  \end{equation}
Here we introduced a condensed notation
$\xi_i=({ u}_i,{ v}_i,{\epsilon}_i)$, so $g(\xi)$ is a distribution of
$u$, $v$, and $\epsilon$,
the $\delta-$function ${\delta}(\xi-\xi_1)={\delta}({ u}-{ u}_1)
{\delta}({ v}-{ v}_1){\delta}({\epsilon}-{\epsilon}_1)$, {\it etc.}
The brackets ${\langle}...{\rangle}_{{ u},{ v},\xi_1,\xi_2}$
stand for averaging ${\int}...
f({ u},{ v})d^2{ u}d^2{ v}g(\xi_1)d^5\xi_1g(\xi_2)d^5\xi_2$.
(Here the integration volume elements are 2- and 5-dimensional because
${ u}$ and ${ v}$ are complex.) The expression
${\langle}...{\rangle}_{{ u},{ v},\xi_1,\xi_2}$ in (\ref{f-rg}) is of
order of $e^{-3 t}$, so the RHS of (\ref{f-rg}) has no explicit dependence on $t$.

The RG flow (\ref{f-rg}) restricted to the distribution of
couplings $ u$, $ v$ is analogous to that discuss in
Sec.~\ref{RG}. One can prove an H-theorem, verify that second
moments are conserved, and derive stationary fix-point
distributions:
  \begin{equation}
  \label{f-fix}
f_{w,z}({ u},{ v})={\pi}^{-2}(w^2-|z|^2)\exp(-w(|{ u}|^2+|{ v}|^2)-
(z{ u}{ v}+c.c.))
  \end{equation}
where $w$ is real and $z$ is complex, satisfying $|z|<w$ for
normalizability of $f_{w,z}(u,v)$.

%%%%%%%%%%%%%%%%%%%%%%%%%%%%%%%%%%%%%%%%%%%%%%%%%%%%%%%%%%%%%%%%%%%%
\begin{figure}
\centerline{\resizebox{10cm}{8cm}{\includegraphics{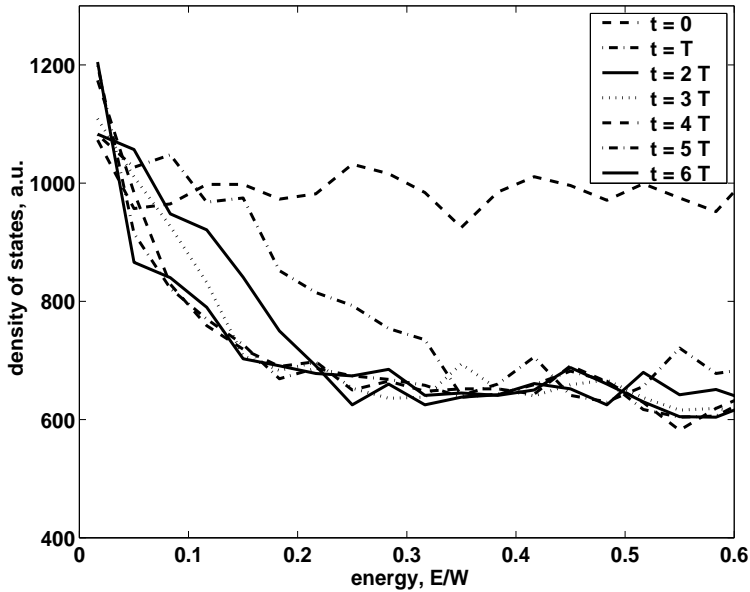}}}
 \caption[]{The flow of the density of states for the distribution of $u$, $v$
    of the form (\protect{\ref{f-fix}}) with $\langle |u|^2\rangle=
    \langle |v|^2\rangle=0.45$, $\langle uv^*\rangle=0$.
    }
 \label{fig5}
\end{figure}
%%%%%%%%%%%%%%%%%%%%%%%%%%%%%%%%%%%%%%%%%%%%%%%%%%%%%%%%%%%%%%%%%%%%%%%

A new aspect of the pseudofermion problem is a nontrivial RG
flow of the density of states. To study how the distribution of
energies ${\epsilon}$ evolves, we consider a particular class of
distributions:
  \begin{equation}
g({ u},{ v},{\epsilon})={\nu}({\epsilon})f_{w,z}({ u},{ v})
  \end{equation}
with $f_{w,z}({ u},{ v})$ being the fix-point
distribution of the form (\ref{f-fix}). (Such an ansatz is
consistent with Eq.(\ref{f-rg}).) Substituting this into
(\ref{f-rg}) one obtains
   \begin{equation}
{{\partial}\over{{\partial}{ t}}}{\nu}({\epsilon})=2{\pi}ne^{3{ t}}
{\langle}{\delta}({\epsilon}-{\epsilon}'_1)+
{\delta}({\epsilon}-{\epsilon}'_2)
-{\delta}({\epsilon}-{\epsilon}_1)-{\delta}({\epsilon}-
{\epsilon}_2){\rangle} _{{ u},{ v},{\epsilon}_1,{\epsilon}_2}
  \end{equation}
We studied the flow of $\nu(\epsilon)$ numerically (see
Figure~\ref{fig5}). There are two distinct features in the flow
of $\nu(\epsilon)$. Initially, at small $t$, the density of
states decreases at all $\epsilon$ due to energy repulsion. At
large $t$, it increases in the vicinity of $\epsilon=0$ forming
a singularity. It is clear from Figure~\ref{fig5} that the first
effect is transitional and quickly saturates, whereas the second
one is characteristic for the scaling limit.

It is interesting to compare this result with the
Efros-Shklovskii theory of Coulomb gap which predicts a
logarithmic suppression of the density of states near
$\epsilon=0$ for the problem (\ref{TLS}) with $z-z$ couplings.
In the above analysis we ignored these couplings because they do
not lead to delocalization of excitations. The effect
of $z-z$ couplings on the density of states is of the same order
as the effects we consider. However, it has an opposite sign.

Therefore, there are two different phases possible in the
problem (\ref{TLS}) depending on relative strength of different
couplings. If the $z-z$ coupling dominates, the system will have
a soft gap in the density of states. On the other hand, if the
$x-x$, $y-y$, and $x-y$ couplings dominate, there will be a peak
in the density of states at $\epsilon=0$. Our analysis indicates
that the peak is described by a power law singularity with the
exponent of order of $\lambda$.

In both phases, however, excitations are delocalized. According
to Secs.~\ref{RG},\ref{Scaling}, the states in this problem
are critical. Delocalized states will give rise to thermal conductivity and,
in the Coulomb gap state, will make variable range hopping conductivity
possible even in the absence of electron--phonon coupling.

\section{Summary}
  \label{Summary}
The RG approach predicts nontrivial fix points of
Hamiltonians with long range hopping. Wavefunctions in such
systems are critical, and the participation ratios are
characterized by scaling exponents which depend on the nature of
the fix point. In some cases, for a particular Hamiltonian different
states can have different exponents, and so the system is
characterized by a distribution of exponents, rather than by a
single exponent. (The possibility of the absence of self-averaging of 
the participation ratio exponent was considered recently \cite{Parshin}.)

The RG method can be applied to the problem of interacting
two-level systems. In doing this it is important to extend the
RG by including renormalization of the density of states.
Depending on the choice of couplings, two different phases are
possible, both characterized by a singularity in the density of
states at $\epsilon=0$. In one phase the density of states
diverges at $\epsilon=0$, whereas in the other phase it
vanishes. It is possible that the first behavior is relevant for
the problem of low energy excitations in glasses, where an
increase in the density of states at low energies due to
interactions between two-level systems has been
conjectured\cite{Kagan}.

%%%%%%%%%%%%%%%%%%%%%%%%%%%%%%%%%%%%%%%%%%%%%%%%%%%%%%%%%%%%%%%%%%%%%%%%%%
%%%%%%%%%%%%%%%%%%%%%% Acknowledgments %%%%%%%%%%%%%%%%%%%%%%%%%%%%%%%%%%
%%%%%%%%%%%%%%%%%%%%%%%%%%%%%%%%%%%%%%%%%%%%%%%%%%%%%%%%%%%%%%%%%%%%%%%%%%
\vspace*{0.25cm} \baselineskip=10pt{\small \noindent
I benefited from discussions with
B.~L.~Altshuler, V.~E.~Kravtsov, A.~D.~Mirlin, and B.~I.~Shklovskii.
This research is supported in part by the MRSEC Program of NSF under 
award 6743000 IRG.}
%%%%%%%%%%%%%%%%%%%%%%%%%%%%%%%%%%%%%%%%%%%%%%%%%%%%%%%%%%%%%%%%%%%%%%%%%%
%%%%%%%%%%%%%%%%%%%%%%%%% Bibliography %%%%%%%%%%%%%%%%%%%%%%%%%%%%%%%%%%%
%%%%%%%%%%%%%%%%%%%%%%%%%%%%%%%%%%%%%%%%%%%%%%%%%%%%%%%%%%%%%%%%%%%%%%%%%%


\begin{thebibliography}{9}

\bibitem{Anderson}
P.~W.~Anderson, Phys. Rev. {\bf 109} (1958) 1492; \\
D. E. Logan, P. G. Wolynes, J. Chem. Phys. {\bf 87} (1987) 7199

\bibitem{LL}
L. S. Levitov,
Phys. Rev. Lett. {\bf 64} (1990) 547;\\
L. S. Levitov,
Europhys. Lett. {\bf 9}(1) (1989) 83

\bibitem{LL1}
B. L. Altshuler, L. S. Levitov,
Physics Reports {\bf 288} (1997) 487

\bibitem{MirlinFyodorov}
Y.~V.~Fyodorov, A.~D.~Mirlin,
Phys. Rev. {\bf B52} (1995) R11580;\\
A.~D.~Mirlin, et al., Phys. Rev. {\bf E54} (1996) 3221

\bibitem{Kravtsov} V. E. Kravtsov, K. A. Muttalib,
Phys. Rev. Lett. {\bf 79} (1997) 1913

\bibitem{Burin} A.~L.~Burin and Yu.~M.~Kagan, Sov. Phys. JETP {\bf 82}
(1996) 159

\bibitem{EfShk}
A. L. Efros, B. I. Shklovskii, J. Phys. {\bf C8} (1975) L49 \\
S. D. Baranovskii, B. I. Shklovskii and A. L. Efros,
Zh. Eksp. Teor. Fiz. {\bf 78} (1980) 395
[Sov. Phys. JETP {\bf 51} (1980) 199]

\bibitem{Yu}
C. C. Yu, A. J. Leggett, Comments Cond. Mat. Phys. {\bf 14} (1988)
231 \\
C. C. Yu, Phys. Rev. Lett. {\bf 63} (1989) 1160

\bibitem{BhattLee}
R. N. Bhatt, P. A. Lee, Phys. Rev. Lett. {\bf 48} (1982) 344

\bibitem{Kagan}
A.~L.~Burin, Yu.~Kagan, L.~A.~Maksimov, I.~Ya.~Polyshchuk,
Phys. Rev. Lett. {\bf 80} (1998) 2945;
Physica. B, Cond. Matt. {\bf 244} (1998) 180;

\bibitem{Parshin}
D.~A.~Parshin, H.~R.~Schober,
Distribution of fractal dimensions at the Anderson transition,
preprint cond-mat/9907067





\end{thebibliography}
\end{document}